\documentstyle[psfig]{mn}

\def\spose#1{\hbox to 0pt{#1\hss}}
\def\approxlt{\mathrel{\spose{\lower 3pt\hbox{$\sim$}} \raise
2.0pt\hbox{$<$}}} \def\approxgt{\mathrel{\spose{\lower
3pt\hbox{$\sim$}} \raise 2.0pt\hbox{$>$}}}

\begin{document}

\title[A Complete Relativistic Ionized Accretion Disc in Cygnus~X--1]
{A Complete Relativistic Ionized Accretion Disc in Cygnus~X--1}
\author[Young et al.]  {\parbox[]{6.in}{A.J. Young$^{1,2}$, A.C.
    Fabian$^1$, R.R. Ross$^3$ and Y. Tanaka$^4$\\ \footnotesize
    $^1$\emph{Institute of Astronomy, Madingley Road, Cambridge CB3
      0HA} \\ $^2$\emph{Present address Department of Astronomy,
      University of Maryland, College Park, MD 20742, USA} \\
    $^3$\emph{Department of Physics, College of the Holy
      Cross, Worcester, MA 01610, USA} \\
    $^4$\emph{Max-Planck-Institut f\"ur extraterrestriche Physik,
      D-85740 Garching, Germany} \\}} \maketitle

\begin{abstract}
  The galactic black hole candidate Cygnus~X--1 is observed to be in
  one of two X-ray spectral states; either the low/hard (low soft
  X-ray flux and a flat power law tail) or high/soft (high
  blackbody-like soft X-ray flux and a steep power law tail) state.
  The physical origin of these two states is unclear. We present here
  a model of an ionized accretion disc, the spectrum of which is
  blurred by relativistic effects, and fit it to the \emph{ASCA},
  \emph{Ginga} and \emph{EXOSAT} data of Cygnus~X--1 in both spectral
  states. We confirm that relativistic blurring provides a much better
  fit to the low/hard state data and, contrary to some previous
  results, find the data of both states to be consistent with an
  ionized thin accretion disc with a reflected fraction of unity
  extending to the innermost stable circular orbit around the black
  hole. Our model is an alternative to those which, in the low/hard
  state, require the accretion disc to be truncated at a few tens of
  Schwarzschild radii, within which there is a Thomson-thin, hot
  accretion flow.  We suggest a mechanism that may cause the changes
  in spectral state.
\end{abstract}


\section{Introduction}
Cygnus X--1 (Cyg~X--1) is one of the brightest and best studied X-ray
sources in the sky. It was the first source to be identified with a
binary system in which the X-ray emission arises through accretion
onto a compact object. (See e.g. Tanaka \& Lewin (1995) for a review
of black hole binaries). It is at a distance of $\sim2$~kpc (Massey et
al. 1995) and consists of a supergiant star and a compact object, with
an orbital period of 5.6 days. The mass of the unseen companion is
significantly greater than $5M_\odot$ (Dolan 1992) suggesting that it
is a black hole. The high energy emission results from focused wind
accretion from the supergiant (Gies \& Bolton 1986) that is almost
filling its Roche lobe. The X-ray flux varies on all timescales from
milliseconds to months although most of its time is spent in the
so-called low or hard state (the `low/hard' state) characterized by a
low soft X-ray flux and strong hard X-ray flux. Occasionally Cyg~X--1
makes a transition from the low/hard state to the high/soft state
which is characterized by a larger soft X-ray flux (by a factor of
$\sim10$) and weaker hard X-ray flux.

The spectrum in the low/hard state above a few keV is typically a
power law of photon index $\Gamma\sim$~1.6--1.8 and a Compton
reflection component above about 10 keV (Done et al. 1992; Ebisawa et
al 1996) with a high-energy cut-off above $\sim 100$ keV. There is
evidence of complicated iron absorption and emission features around
6--7 keV (Barr, White \& Page 1985; Marshall et al. 1993; Ebisawa et
al.  1996).

In the high/soft state the spectrum is dominated by a soft
blackbody-like component peaking around 1 keV with a power law tail
of photon index $\Gamma\approx2.5$ (Gierli\'nski et al. 1999).

It has been suggested that the different spectral states are due to a
qualitative change in the accretion flow (see e.g. Gierli\'nski et al.
1997; Esin et al. 1998; Poutanen \& Coppi 1998; Done \& \.Zycki 1999).
In the high/soft state there is a `standard' accretion disc extending
in to the innermost stable circular orbit around the black hole, and
in the low/hard state the region within a few 10s of Schwarzschild
radii becomes a geometrically thick, optically thin, radiatively
inefficient advective flow (see also Nowak et al. 1999). The evidence
for this is that in the low/hard state the reflected fraction appears
to be low, which is consistent with the innermost parts of the
accretion disc being missing, or at least Thomson thin. The change
between spectral states occurs with a small change in bolometric
luminosity ($\sim50$ per cent) and no change in $1.3-200$ keV
luminosity to within $\sim15$ per cent (Zhang et al. 1997).

In this paper we investigate an alternative model in which the low
inferred reflected fraction is a result of the ionization state of the
accretion disc. A highly ionized disc may be an almost perfect
reflector with weak reflection features (as compared to cold
reflection) and hence may lead to an underestimate of the reflected
fraction (Ross, Fabian \& Young 1999; Ross, Fabian \& Brandt 1996;
Ross \& Fabian 1993; Nayakshin, Kazanas \& Kallman 2000). The
attraction of this model is that it does not require the nature of the
accretion flow to change qualitatively between the spectral states; a
change which is observed to occur with little variation in bolometric
luminosity as observed. It is just a change in density at the surface
of the disc.

Previous conclusions that the disc cannot be significantly ionized
have been based on the absence of a strong photoelectric absorption
edge at 8 keV. In the case of an ionized disc absorption and emission
may occur at significant Thomson depths ($\tau_{\rm T}\approxgt1$) and
sharp features are blurred due to Compton scattering. See e.g. Fig.~7
of Ross, Fabian \& Young (1999).

In view of the extremely rapid continuum variability of Cyg~X--1 we
may also expect its spectrum to change rapidly. The spectra that we
fit are necessarily time averaged since a typical observation will
last for millions of dynamical timescales.

\section{Data and Model}
\subsection{Data}
We use \emph{ASCA} SIS, \emph{ASCA} GIS, \emph{EXOSAT} and
\emph{Ginga} data of Cyg~X--1 in the low/hard state and \emph{ASCA}
SIS data of Cyg~X--1 in the high/soft state. We do not use the
\emph{ASCA} 93~Oct data of the low/hard state since this was taken
when the satellite was in daytime (Ebisawa et al. 1996). The data used
are summarized in Table~\ref{tab_data}.

\begin{table}
\begin{tabular}{lllll}
No. & Date & Observation & State \\
\hline
 1 & 1993 Oct & \emph{ASCA} GIS & Low/hard \\
 3 & 1993 Nov 11 & \emph{ASCA} GIS & Low/hard \\
 4 & 1993 Nov 12 & \emph{ASCA} GIS & Low/hard \\
 5 & 1993 Nov & \emph{ASCA} SIS & Low/hard \\
 6 & 1994 Nov & \emph{ASCA} GIS, no.1 & Low/hard \\
 7 & 1994 Nov & \emph{ASCA} GIS, no.2 & Low/hard \\
 8 & 1994 Nov & \emph{ASCA} GIS, no.3 & Low/hard \\
 9 & 1994 Nov & \emph{ASCA} GIS, no.4 & Low/hard \\
10 & 1994 Nov & \emph{ASCA} GIS, no.5 & Low/hard \\
 a & 1991 Jun 6 & \emph{Ginga} no. 2 & Low/hard \\
 b & 1996 May & \emph{ASCA} SIS & High/soft \\
E08 & 1984 Jul 9 & \emph{EXOSAT} GSPC & Low/hard \\
E09 & 1985 Sep 14 & \emph{EXOSAT} GSPC & Low/hard \\
E10 & 1985 Aug 12 & \emph{EXOSAT} GSPC & Low/hard \\
E13 & 1984 Nov 2 & \emph{EXOSAT} GSPC & Low/hard \\
E14 & 1984 Nov 2 & \emph{EXOSAT} GSPC & Low/hard \\
\hline
\end{tabular}
\\
\caption{Data sets used. The data set numbers 1 and 3--10 correspond
  to those of Ebisawa et al. (1996), and those prefixed by `E' correspond
  to those of Done et al. (1992).}
\label{tab_data}
\end{table}

\subsection{Rest frame reflection spectra}
\begin{figure}
\centerline{\psfig{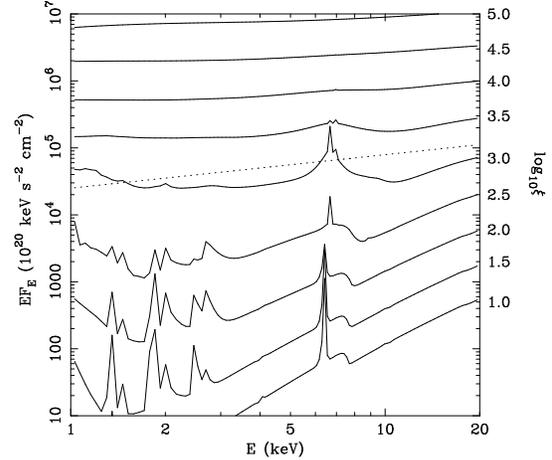}}
\caption{Reflection spectra from a slab of gas illuminated by an X-ray
power law of photon index $\Gamma=1.6$, solar iron abundance and
various values of the ionization parameter $\xi$. The lowest solid
line is for $\log \xi=1$, and $\log \xi$ increases upwards to $\log
\xi=5$ in steps of 0.5. The dotted line is a representative
illuminating power law, in this instance the $\log \xi=3$ case.}
\label{rs}
\end{figure}

The accretion disc model is based on the X-ray reflection spectra from
ionized slabs of gas calculated by Ross, Fabian \& Young (1999) and
Ross \& Fabian (1993). The reflection spectrum resulting from the
illumination of a slab of gas by an X-ray power law of photon index
$\Gamma$ depends primarily on the ionization parameter $\xi$, defined
as the ratio of flux and density,
\[ \xi=\frac{4\pi F_{\rm x}}{n_{\rm H}} \]
where $F_{\rm x}$ is the illuminating X-ray flux (0.01--100 keV) and
$n_{\rm H}$ is the hydrogen number density. Fig.~\ref{rs} shows
reflection spectra calculated for solar iron abundance, a photon index
of 1.6 and a range of ionization parameters, $\log\xi=1-5$ in steps of
0.5.

For very low ionization parameters the most prominent feature of the
reflected spectrum is the iron fluorescence line at 6.4~keV. In
neutral gas both oxygen and iron are strong absorbers but the
absorption above and below the iron edge is so strong that the change
in observed flux across the edge is small, particularly if the primary
continuum is also observed. This absorption also tends to harden the
spectrum as compared to the illuminating power-law.

As the X-ray illumination of the slab becomes more intense the matter
becomes more highly ionized. This reduces the effects of absorption
progressively from lower to higher energies. An important signature of
ionized reflection is a strong iron edge. As much of the oxygen
becomes completely ionized the edge becomes more prominent as the
absorption below it is reduced. As the outer layers of the slab become
completely ionized fluorescent photons are produced deeper in the slab
and are Compton scattered on leaving the slab resulting in broad and
blended emission and absorption features, e.g. the case of $\xi=10^3$
in Fig.~\ref{rs}. The narrow line cores are from those line photons
that escape without scattering or are generated closer to the
surface. For an extremely highly ionization parameter $\xi=10^5$ the
reflection spectrum is almost indistinguishable from the illuminating
continuum.

\subsection{Simple accretion disc model}
Initially we consider a model of a non-relativistic accretion disc
with a single ionization parameter $\xi$. This consists of two
components; the primary X-ray source which we assume to be a power law
of photon index $\Gamma$ and the reflected spectrum corresponding to
some fraction of these X-rays being Compton scattered by the disc into
our line of sight.

We have computed a grid of reflection spectra for a range of photon
indices, ionization parameters and iron abundances, and subsequently
converted to an {\sc xspec} `tablemodel'. {\sc xspec} is the spectral
fitting code commonly used by X-ray astronomers. The reflection
spectra in the tablemodel have been scaled so that, in {\sc xspec}, a
reflection spectrum plus power law of the same photon index has a
corresponding reflected fraction equal to the ratio of the
normalization of the reflected spectrum to the power law. We have
considered ionization parameters in the range $\xi=10^1$--$10^5$ (up
to $\xi=10^6$ in the high/soft case) and iron abundances in the range
1--$3\times$solar.

\subsection{Relativistic blurring}
The accretion disc is deep in the potential well of the black hole and
subject to strong Doppler, transverse Doppler and gravitational
redshifting effects. These will smear out the sharp features in the
spectrum and blend the emission and absorption features together. The
detection of such an effect in the spectrum of Cyg~X--1 in the
low/hard state has been reported by Done \& \.Zycki (1999) but with
quite different parameters to those reported here. We compute this in
{\sc xspec} by convolving the spectrum with a blurring kernel taken
from the {\sc diskline} model for an accretion disc around a
Schwarzschild black hole (Fabian et al. 1989). Special care needs to
be taken when using this approach in {\sc xspec} since it requires the
model to be evaluated outside the energy range of the data. This is
discussed in Appendix~A.

The parameters of the blurring model define the inner and outer radii
of our accretion disc. The outer radius is fixed at $1000 r_g=1000
GM/c^2$, and the inner radius $r_{\rm in}$ must be $\ge 6 r_g$ ($6 r_g
= 6 GM/c^2 = 3 R_S$, where $R_S$ is the Schwarzschild radius, and is
the location of the innermost stable circular orbit around a
Schqarzschild black hole). The precise value of the outer radius has
little impact on the results we present here.  The `radial emissivity
profile' parameter required for the {\sc diskline} kernel is set to
$(1-\sqrt{6/r_g})/r_g^3$.

\subsection{A more realistic disc model}\label{sec:model}
We shall find that some of the data sets require the use of a more
sophisticated model. In the future it will be useful to construct a
proper model that accounts for the radial variation of ionization
parameter in more detail, but for the time being we use the
prescription described below.

The accretion disc is divided into two zones, an `inner' region which
will have a relatively high ionization parameter, and an `outer'
region with a lower ionization parameter. The inner and outer regions
are convolved with {\sc diskline} kernels extending between
$6r_g\rightarrow r_{\rm trans}$ and $r_{\rm trans}\rightarrow1000 r_g$
respectively, where $r_{\rm trans}$ represents the radius of the
transition between high and low ionization zones. The relative
contribution to the reflection from regions $r < r_{\rm trans}$ and $r
> r_{\rm trans}$ are calculated to agree with the Page \& Thorne
(1974) radial emissivity law. From a practical point of view, fitting
the data using {\sc xspec}, this achieved iteratively, i) fitting the
data, ii) fixing $r_{\rm trans}$ at the appropriate value given the
relative contribution to the reflection from the two regions, iii)
repeating until an self-consistent fit is achieved. Very few
itierations were required. The error in determining $r_{\rm trans}$ is
quite small as an uncertainty of $\sim 10$ per cent in the ratio of
fluxes would correspond to an error of $\pm 2 r_g$ in $r_{\rm trans}$.
This represents a realistic model of an accretion disc with a
\emph{sharp} transition from high to low ionization state.  Whether
such an abrupt transition is \emph{required} will be investigated
later.

\section{Fits to \emph{ASCA} SIS Data}
In this section we consider various fits to the \emph{ASCA} SIS data
of Cyg~X--1 in both the high/soft and low/hard states. We are mainly
interested in fitting the data between 4--10 keV since this is where
most of the ionized reflection features are expected to lie. It is
also uncertain as to whether the primary X-ray continuum is actually a
power-law over the entire \emph{ASCA} energy range.

\subsection{Simple disc model}
\subsubsection{Low/hard state}
\begin{figure}
\centerline{\psfig{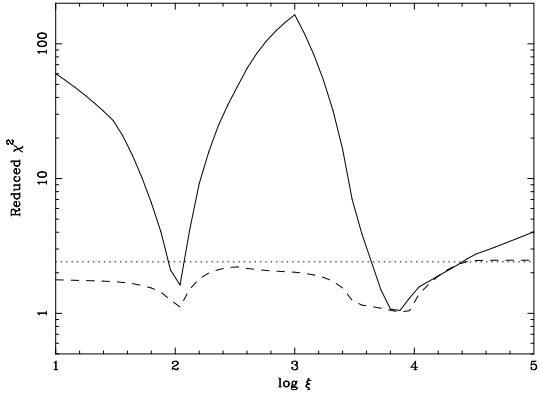}}
\caption{(Low/hard). Results of fitting the ionized disc model,
without relativistic blurring or additional spectral components, to
the 93 Nov \emph{ASCA} data of Cyg~X--1 in the low/hard state. The
solid curve shows the reduced $\chi^2$ value for a reflected fraction
of 1, the dotted line for a reflected fraction of 0 and the dashed
line for the reflected fraction being a free parameter. When the
reflected fraction is a free parameter it has been fitted for at each
value of $\xi$.}
\label{lh1}
\end{figure}

We begin by fitting a non-relativistically blurred ionized disc model
to the 4--10 keV \emph{ASCA} SIS data of the low/hard state. In these
fits the iron abundance has been fixed at twice solar, the best
fitting value suggested by Done \& \.Zycki (1999).  Such a model is
not able to give a good fit to the data but it is interesting to
consider a plot of $\chi^2$ against $\xi$ shown in Fig.~\ref{lh1}. The
solid curve shows the quality of fit with the reflected fraction $f$
frozen at 1, i.e. with the disc subtending $2\pi$~sr of the sky as
seen by the X-ray source. There are two local minima, one at
$\xi=10^2$ and the other around $\xi=10^4$. The best fitting solution
obtained will depend upon the initial value of $\xi$, the separatrix
being at approximately $\xi=10^3$. The horizontal dotted line shows
the $\chi^2$ value for a reflected fraction of 0, and the dashed line
shows the values with the reflected fraction free parameter.

In the case of a low ionization parameter $\xi=10$ a fit with a
reflected fraction of 1 is significantly worse than that of just a
power law. A much smaller reflected fraction $f=0.07$ provides a
better fit. For an ionization parameter of $\xi=10^2$ at the first
minima in $\chi^2$ we again find a low reflected fraction is
preferred, with $f=0.55$. At the second minima $\xi=10^{3.8}$,
however, we find the best fitting reflected fraction to be
$f=1.07$. Table~\ref{tab1} summarizes the results of these provisional
fits. The values of the photon index $\Gamma$ are consistent with the
results of Gierli\'nski et al. (1997).

\begin{table}
\begin{tabular}{lllllr}
Data set & State & $\log \xi$ & $\Gamma$ & $^bf$ & $\chi^2/{\rm dof}$
\\ \hline
5 & low/hard  & $1.0^a$ & 1.57 & 0.07 & 128/76 \\
5 & low/hard & $2.0^a$ & 1.65 & 0.55 &  78/76 \\
5 & low/hard  & $3.8^a$ & 1.68 & 1.07 &  75/76 \\
5 & low/hard  & $5.0^a$ & 1.54 & 0.00 & 178/76 \\
b & high/soft & $1.0^a$ & 2.36 & 0.17 &  57/37 \\
b & high/soft & $2.4^a$ & 2.47 & 1.14 &  33/37 \\
b & high/soft & $5.1^a$ & 2.32 & 1.13 &  37/37 \\
b & high/soft & $6.0^a$ & 2.21 & 162 &  70/37 \\ \hline
\end{tabular}
\\
\begin{flushleft}

$^a$ frozen \\ $^b$ reflected fraction ($2\pi$ sr) \\

\end{flushleft}
\caption{Fit statistics for an ionized accretion disc model, without
relativistic blurring or additional spectral components. Although some
of these fits appear to be statistically acceptable some show
systematic residuals at the expected energies of line and edge
features.}
\label{tab1}
\end{table}

To summarize, we have found two distinct solutions, one preferring a
low ionization parameter and low reflected fraction, the other
preferring high ionization parameter and a high reflected fraction. We
shall investigate both of these with a more realistic physical model.

\subsubsection{High/soft state}
\begin{figure}
\centerline{\psfig{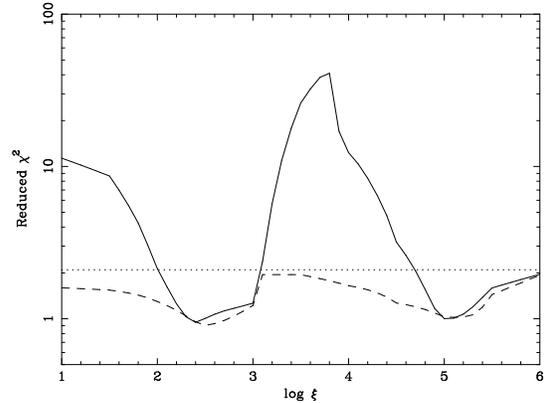}}
\caption{(High/soft). Results of fitting the ionized disc model,
without relativistic blurring or additional spectral components, to
the \emph{ASCA} data of Cyg~X--1 in the high/soft state. The solid
curve shows the reduced $\chi^2$ value for a reflected fraction of 1,
the dotted line for a reflected fraction of 0 and the dashed line for
the reflected fraction being a free parameter. When the reflected
fraction is a free parameter it has been fitted for at each value of
$\xi$.}
\label{hs1}
\end{figure}

Similar results are obtained for the high/soft state, where we have
considered ionization parameters up to $\xi=10^6$, as shown in
Fig.~\ref{hs1}. The location of the minima have changed to
$\xi=10^{2.5}$ and $\xi=10^5$. Table~\ref{tab1} also shows the results
of these provisional fits.  A low reflected fraction seems to be
preferred for an unionized disc although a fraction closer to 1 is
acceptable in the case of the disc being ionized. The values of the
photon indices $\Gamma$ are consistent with those found by
Gierli\'nski et al. (1999) when they include \emph{OSSE} $\gamma$-ray
observations.

\subsection{Relativistic blurring}
\subsubsection{Low/hard state}
\begin{figure}
\centerline{\psfig{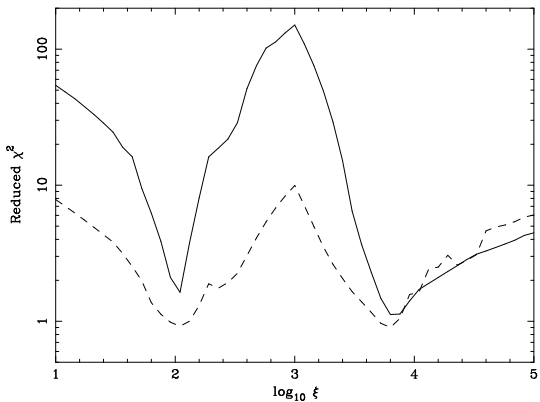}}
\caption{(Low/hard). Results of fitting the relativistically blurred
  ionized disc model to the 93 Nov \emph{ASCA} data of Cyg~X--1 in the
  low/hard state. The reflected fraction has been fixed at 1. The
  solid curve shows the fit without relativistic blurring, and the
  dashed line shows the improvement with blurring from an accretion
  disc spanning $6-50r_g$. The inclination angle is a free parameter,
  fitted for at each value of $\xi$.}
\label{lhrb}
\end{figure}

We fit the \emph{ASCA} data of the low/hard state between 4--10 keV
with the ionized disc model convolved with the relativistic blurring
model described above. The reflected fraction is fixed to 1 and the
iron abundance to twice solar. Without relativistic blurring we found
there to be two local minima in $\chi^2$ around $\xi=10^2$ and
$\xi=10^{3.8}$. We first consider an accretion disc with the inner
radius fixed at $r_{\rm in}=6 r_g$, the outer radius fixed at $50 r_g$
and the inclination a free parameter. This improves the quality of fit
significantly around $\xi=10^2$ and only slightly around
$\xi=10^{3.8}$ to such an extent that the $\xi=10^2$ solution is now
(only just) favoured. If we now allow the inner disc radius to be a
free parameter we get a better fit for very low values of the
ionization parameter but almost no change at the best fitting
values. These results are shown in Fig.~\ref{lhrb}.

\subsubsection{High/soft state}
\begin{figure}
\centerline{\psfig{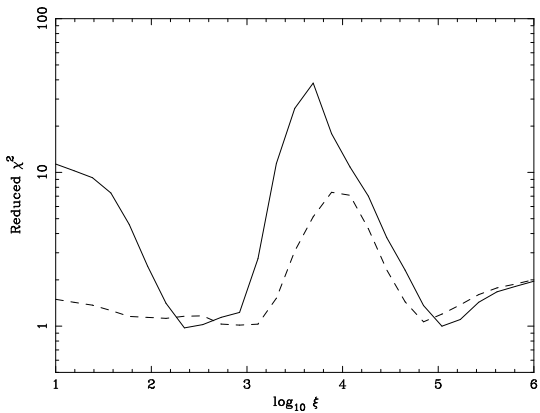}}
\caption{(High/Soft). Results of fitting the relativistically blurred
  ionized disc model to the 93 Nov \emph{ASCA} data of Cyg~X--1 in the
  high/soft state. The reflected fraction has been fixed at 1. The
  solid curve shows the fit without relativistic blurring, and the
  dashed curve shows the change when blurring from an accretion disc
  spanning $6-50r_g$ is included.}
\label{hsrb}
\end{figure}

In contrast to the low/hard state Fig.~\ref{hsrb} shows that
relativistic blurring does not significantly improve the quality of
the best fits, and it is possible to achieve an equally good fit with
or without blurring, with ionization parameters of $\xi=10^{2.5}$ and
$\xi=10^{5}$. In the case of cold reflection, however, blurring does
significantly improve the quality of fit.

\section{Fits to the \emph{ASCA}, \emph{Ginga} and \emph{EXOSAT} data}

When we fit the \emph{ASCA} GIS data we find that a satisfactory fit
cannot be achieved with a disc that has a single ionization parameter.
This is not too surprising since the the models is extremely
simplified, and the GIS data have a better spectral resolution that
the SIS data at higher energies, and provide a more stringent test for
theoretical models. A model in which the disc is divided into two
regions, each with a different ionization parameter, does provide a
satisfactory fit to the data. This is not a perfect model, but it is
more realistic than one in which there is a single ionization
parameter. The details of the model are discussed in
section~\ref{sec:model}. To fit the data with such a model we
initially allow $r_{\rm trans}$ and the relative normalizations of the
inner and outer components to be free parameters. Fitting this model
to the data gives the approximate relative normalization of the inner
and outer components allowing the corresponding $r_{\rm trans}$ to be
computed. $r_{\rm trans}$ is then frozen at this value and the data
are fit again. We check the relative fluxes of the two components are
still consistent with the $r_{\rm trans}$. This ensures that each of
the components is blurred appropriately.

The results of fitting this model to the \emph{ASCA} SIS and GIS,
\emph{Ginga} and \emph{EXOSAT} data are summarized in
table~\ref{tab:2xifit}. The reduced $\chi^2$ values of the fits to the
\emph{ASCA} GIS data are comparable to those found by Ebisawa et al.
(1996) and Done \& \.Zycki (1999). Overall this model is able to
provide a good fit to the available data. The preferred inclination
angles tend to be small which is in reasonable agreement with the
optical studies of Gies \& Bolton (1986) in which the most probable
inclination angle is $28^\circ-38^\circ$.

\begin{table*}
\begin{tabular}{llllllrcl}
Data set & $^a$PL $\Gamma$ & $^b \log_{10}\xi_{\rm inner}$ & $r_{\rm
  trans}$ ($r_g$) & $\log_{10}\xi_{\rm outer}$ & $^c$inc & $\chi^2$
&/& d.o.f. \\
\hline

1 & 1.79 & $4.4^{+0.1}_{-0.2}$ & 26 & $1.8^{+0.1}_{-0.2}$ & 33 & 111 &/&
85 \\

3 & 1.75 & $4.4^{+0.1}_{-0.1}$ & 28 & $2.0^{+0.1}_{-0.0}$ & 11 & 75 &/&
81 \\

4 & 1.61 & $5.0_{-0.2}$ & 22 & $2.0^{+0.2}_{-0.0}$ & 8 & 129 &/&
81 \\

5 & 1.59 & $3.9^{+0.1}_{-0.0}$ & 26 & $2.0^{+0.1}_{-0.1}$ & 7 & 66 &/&
77 \\

6 & 1.57 & $4.4^{+0.1}_{-0.1}$ & 30 & $1.9^{+0.1}_{-0.0}$ & 0 & 100 &/&
84 \\

7 & 1.49 & n/a & 6 & $2.0^{+0.0}_{-0.0}$ & $41^{+5}_{-3}$ & 77 &/& 87 \\

8 & 1.52 & $4.3^{+0.1}_{-0.1}$ & 36 & $2.0^{+0.0}_{-0.2}$ & 4 & 115 &/&
85 \\

9 & 1.50 & $2.3^{+0.4}_{-0.8}$ & 36 & $1.6^{+0.2}_{-0.6}$ & 1 & 92 &/&
85 \\

10 & 1.46 & $5.0^{+}_{-0.3}$ & 22 & $2.0^{+0.1}_{-0.1}$ & 7 & 103 &/& 85
\\

a & 1.58 & $4.5^{+0.5}_{-0.3}$ & 20 & $2.0^{+0.3}_{-0.4}$ & 1 & 4 &/& 13
\\

E08 & 1.65 & $4.4^{+0.4}_{-0.3}$ & 30 & $1.8^{+0.1}_{-0.2}$ & 23 & 57
&/& 80 \\

E09 & 1.50 & $5.0_{-0.7}$ & 12 & $1.9^{+0.1}_{-0.4}$ & 20 & 81 &/&
67 \\

E10 & 1.44 & $4.3^{+0.2}_{-0.1}$ & 43 & $1.7^{+0.2}_{-0.3}$ & 20 & 53
&/& 76 \\

E13 & 1.68 & $4.7^{+0.3}_{-0.3}$ & 21 & $1.7^{+0.2}_{-0.2}$ & 38 & 143
&/& 152 \\

E14 & 1.61 & $4.3^{+0.2}_{-0.1}$ & 24 & $2.0^{+0.0}_{-0.0}$ & 20 & 64
&/& 75 \\

\hline

b & 2.46 & n/a & 6 & $2.2^{+0.2}_{-0.2}$ & 13 & 36 &/& 36 \\

\hline
\label{tab:2xifit}
\end{tabular}
\begin{flushleft}
  
  $^a$ power law photon index $\Gamma$ \\
  
  $^b$ 5.0 is the upper limit for this parameter \\

  $^c$ accretion disc inclination angle

\end{flushleft}
\caption{Fit statistics for our model of a complete ($6 r_g
  \rightarrow 1000 r_g$) relativistically smeared ionized accretion
  disc with a reflected fraction of unity. The low/hard state data are
  above the separating line, the high/soft state data below.}
\end{table*}

\begin{figure}
\centerline{\psfig{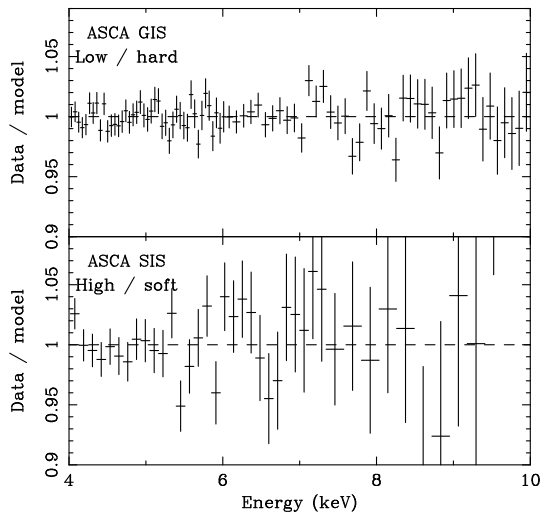}}
\caption{Typical ratio plots of good fits to the \emph{ASCA} data of
Cyg~X--1 in both the low/hard and high/soft state.}
\label{fig:ratio}
\end{figure}

\begin{figure}
\centerline{\psfig{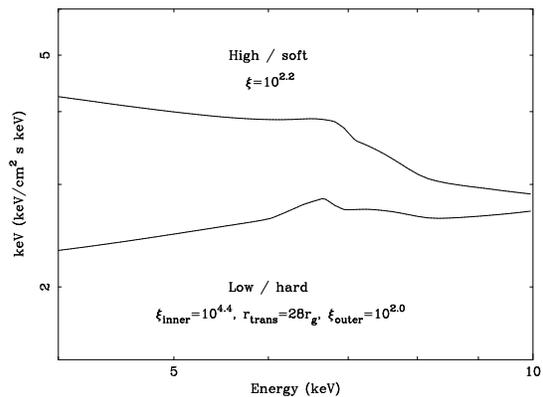}}
\caption{Two models that provide good fits to the \emph{ASCA} data of
Cyg~X--1 in the high/soft and low/hard states.}
\label{fig:models}
\end{figure}

The low/hard state may be characterized by an accretion disk in which
the inner region within $\sim13$ Schwarzschild radii is highly
ionized, $\xi\sim10^4$, and outside of which the disk is less highly
ionized, $\xi\sim10^2$. In the high/soft state we have only fitted one
data set, but that is consistent with a relatively cool accretion disk.
In all cases the accretion disk extends in to the innermost stable
orbit and has a reflection fraction of unity. Fig.~\ref{fig:ratio}
shows typical fits to the \emph{ASCA} GIS data of the low/hard state
and \emph{ASCA} SIS data of the high/soft state. The corresponding
models are shown in Fig.~\ref{fig:models}

Note that in these fits we have not added any contribution to the
spectrum from thermal disk emission (which will have a small effect on
these fits since we are fitting above 4 keV), or any contribution to
the iron line from reflection off the companion star. This would
consist of a narrow iron line at 6.4 keV of low equivalent
width. Either of these measures could improve the quality of fit
listed in table~\ref{tab:2xifit}.

\section{Transition region}

So far we have assumed a sharp transition between the high ionization
inner zone and the lower ionization outer zone. It is important to try
and quantify the sharpness of this transition. Some models such as
those of Ross \& Fabian (1993) predict that as the ionization
parameter decreases outwards there should be a zone in which H-like
and He-like iron fluorescence is observed. This has twice the
fluorescence yield of neutral iron, and produces strong lines at 6.67
keV and 6.97 keV.  Other models, such at those of Nayakshin et al.
(2000) and Nayakshin (2000) predict the strength of the iron line to
decrease monotonically with radius, i.e. it is not necessary to have
an annulus in which strong H-like and He-like iron fluorescence is
produced.

\begin{figure}
\centerline{\psfig{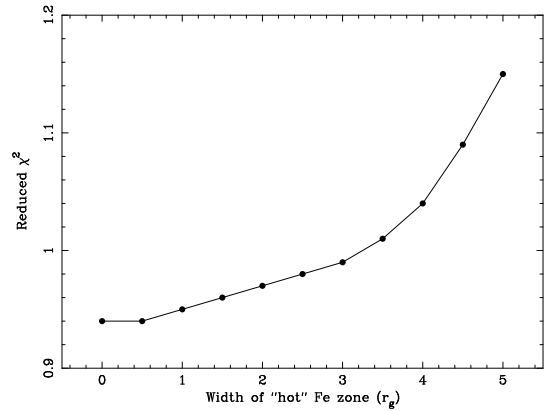}}
\caption{Change in $\chi^2$ of fit to data set 3 (\emph{ASCA} GIS)
  with the introduction of a transition zone of width $\Delta r$ in
  which the reflected spectrum has H-like and He-like iron lines. The
  best overall fit is achieved with a sharp transition (i.e. $\Delta
  r=0$) but the data are not of sufficient quality to rule out a
  gradual transition.}
\label{trans}
\end{figure}

To investigate whether an intermediate zone in which H-like and
He-like iron are produced is consistent with the data we modified the
ionized disc model as follows. An extra zone was introduced of width
$\Delta r$, with an ionization parameter fixed so that the reflected
spectrum had H-like and He-like iron emission, and the normalization
fixed so that the ratio of model flux in $6r_g\rightarrow r_{\rm
  trans}$ to that in $r_{\rm trans}\rightarrow r_{\rm trans}+\Delta r$
is equal to that expected for a standard accretion disc. Fits were
then performed to see how the best fitting $\chi^2$ values changed
with $\Delta r$.  For data set 3 (\emph{ASCA} GIS) the results are
shown in Fig.~\ref{trans}.

The introduction of a zone of H-like and He-like iron makes the fit
progressively worse as the width of the zone increases. The overall
best fit is achieved without such a zone, i.e. with a sharp
transition. With the present quality of data, however, acceptable fits
are obtained with a zone of reasonable width, $\sim$ few $r_g$, and we
cannot rule out a gradual change in ionization parameter.

\section{Discussion}
We have shown that a relativistically blurred accretion disc with two
ionization states is able to fit the X-ray data of Cyg~X--1. The disc
may extend in to the innermost stable orbit around the black hole, and
have a reflected fraction of 1. We present this model as an
alternative to those in which, in the low/hard state, the outer
accretion disc is not ionized and extends in to a few 10s of
Schwarzschild radii, within which the accretion flow becomes
geometrically thick, optically thin and radiatively inefficient. One
attraction of our model is that the accretion flow need not change
qualitatively between the two spectral states of Cyg~X--1; it just
requires a change in the surface density of the disc.
Fig.~\ref{fig:models} shows two representative fits to the data.

It is also clear that, in the low/hard state, relativistic blurring of
the disc spectra results in a significant improvement in the quality
of fit.

It should be noted that our model is simplified in that the accretion
disc is assumed to have two ionization zones. With the present data we
are unable to investigate the radial (or indeed azimuthal) dependence
of the ionization parameter in sufficient detail.

\subsection{State changes in Cyg X--1}
We now discuss a possible accretion disc model to explain the
differing spectral properties of Cyg~X--1 in the low/hard and
high/soft states. Most of the time Cyg~X--1 is in the low/hard state
and it makes occasional transitions to the high/soft state. The
mechanism for this transition is at present poorly understood. We
propose a model in which the state change is due to a change in the
ratio, $f_{\sc c}$, of accretion power dissipated in the corona to
that dissipated in the disc.

If most of the accretion power is dissipated in the disc, $f_{\rm
  c}\approxlt 0.1$, we expect significant thermal blackbody emission
from the disc. There are then many UV seed photons to be inverse
Compton scattered in the corona and the X-ray spectrum will be
relatively soft. If on the other hand significant power is dissipated
in the corona, $f_{\rm c}\approxgt 0.4$, we expect the blackbody
emission to be reduced. The seed-photon-starved corona will produce a
harder spectrum. A combination of factors may then give rise to the
observed behaviour of the ionization parameter. Firstly, a larger
fraction of accretion power being dissipated in the corona provides
increased hard X-ray illumination of the disc, and hence a larger
value of $\xi$ (note that the bolometric luminosity remains constant
with changing $f_{\rm c}$ but the ratio of hard to soft flux does
change). Secondly, the inverse Compton temperature is much higher for
a hard illuminating spectrum than a soft one, and if the outermost
layers of the disc heat up to the Compton temperature (e.g. see
R\'o\.za\'nska 1999; Nayakshin, Kazanas \& Kallman 2000) this will be
important in determining the oberved value of $\xi$. A final
possibility is that the density of the surface layers of the disc
changes between spectral states, with the low/hard state having a
lower surface density (giving rise to a higher value of $\xi$) than
the high/soft state. Table~\ref{tab3} summarizes the expected
properties of this simple model.

\begin{table}
\begin{tabular}{lllll}
State & $^a\xi$ & $^bf_{\rm c}$ & $^cf$ & $^d\Gamma$ \\ \hline
Low/hard & High & $\approxgt0.4$ & Low & Low\\ High/soft & Low &
$\approxlt0.1$ & High & High\\ \hline
\end{tabular}

$^a$ ionization parameter \\ $^b$ fraction of accretion power
dissipated in the corona \\ $^c$ inferred reflected fraction \\ $^d$
observed photon index \\
\caption{Model predictions; see discussion}
\label{tab3}
\end{table}


We also note that the predictions of changing spectral slope are
consistent with the calculations of Coppi (1999) in which an increase
in the soft flux coupled with a reduction of the hard flux leads to an
increase in the photon index of the continuum, i.e. a softer spectrum.

Our simple model is also consistent with the observational results of
Zdziarski, Lubi\'nski \& Smith (1999) whereby the inferred (cold)
reflected fraction is correlated with the observed photon index. The
correlation is almost linear ranging from a reflected fraction of 0
with a photon index of $\Gamma=1.6$ up to a reflected fraction of 2
with a photon index of $\Gamma=2.2$, with some scatter. This
correlation is also seen within individual objects whose photon
indices and reflected fractions change with time. We predict that a
softer spectrum (high photon index) should have a lower value of the
ionization parameter and hence most of the disc may produce a cold
line resulting in a large inferred reflected fraction. In the case of
a harder spectrum (low photon index) we predict a higher value of the
ionization parameter and the spectrum will be that of an ionized disc
as discussed above. This may result in the inferred reflected fraction
being lower. Another model that is able to reproduce this correlation
is that of Beloborodov (1999) in which the X-ray sources move away
from the disc at mildly relativistic speeds.

Further detailed calculations are, of course, required to investigate
this model properly. It is not clear how Cyg~X--1, and other BHC,
remain `locked' into their spectral states when the dynamical
timescale is so short.

\subsection{Future observations and theory}
Future observations with higher spectral resolution will provide much
more stringent tests of the accretion disc models for Cyg~X--1 and
other BHC. At present, even with the \emph{ASCA} data, we are unable
to place strong constraints on the parameter space of our model (e.g.
inner disc radius, reflected fraction) as our statistics are not good
enough.

This model is also applicable to other black hole candidates and we
hope to investigate these in the future. More complete models that
account for the detailed ionization structure of the accretion disc as
a function of radius need to be computed.

\section{Acknowledgements}
We thank Ken Ebisawa for the \emph{ASCA} and \emph{Ginga} data of
Cyg~X--1 in the low/hard state, and Keith Arnaud for advice on solving
our convolution problem. We would also like to thank the referee P. T.
\.Zycki for helpful comments. AJY thanks PPARC for support. ACF thanks
the Royal Society.

\appendix
\section{Blurring model spectra}
The disc spectra are convolved with the kernel of the {\sc diskline}
model (Fabian et al. 1989). In order to do this our disc spectrum must
be evaluated outside the range over which we are fitting the data.
This is because, for example, higher energy photons may be redshifted
into the band we are fitting. If the response matrix of the detector
is such that the set of channels that correspond to a higher energy
photon is distinct from the set of channels channels that correspond
to a photon in the energy range being fitted then the model will not
be evaluated at that higher energy and hence that contribution will be
lost. This leads to the model spectrum dropping off at the edges of
the range being fitted.

A solution to this is to firstly combine the response and auxiliary
response matrices, and to replace the auxiliary response matrix with
one in which the effective area of each energy bin is unity. The {\sc
  extend} command may then be used to enlarge the energy range over
which the model is evaluated. This solves the problem.
\end{document}